\documentclass[12pt,a4paper,final]{iopart}

\usepackage[breaklinks=true, colorlinks=true, linkcolor=blue, urlcolor=blue, citecolor=blue]{hyperref}
\usepackage{amsfonts, amsmath, dsfont}
\usepackage{cite}

\def\epp{\, .}
\def\epc{\, ,}
\newcommand{\sine}[2]{s_{#1,#2}}
\newcommand{\sss}{\scriptscriptstyle}

\begin{document}

\title[N\'eel-XXZ state overlaps: odd particle numbers]{N\'eel-XXZ state overlaps: odd particle numbers and Lieb-Liniger scaling limit}

\author{M.~Brockmann, J.~De Nardis, B.~Wouters, and J.-S.~Caux
}
\address{
Institute for Theoretical Physics, University of Amsterdam, Science Park 904,\\
Postbus 94485, 1090 GL Amsterdam, The Netherlands}
\ead{M.Brockmann@uva.nl}

\begin{abstract}
We specialize a recently-proposed determinant formula \cite{XXZpaper} for the overlap of the zero-momentum N\'eel state with Bethe states of the spin-1/2 XXZ chain to the case of an odd number of downturned spins, showing that it is still of ``Gaudin-like'' form, similar to the case of an even number of down spins. We generalize this result to the overlap of $q$-raised N\'eel states with parity-invariant Bethe states lying in a nonzero magnetization sector. The generalized determinant expression can then be used to derive the corresponding determinants and their prefactors in the scaling limit to the Lieb-Liniger (LL) Bose gas. The odd number of down spins directly translates to an odd number of bosons. 

We furthermore give a proof that the N\'eel state has no overlap with non-parity-invariant Bethe states. This is based on a determinant expression for overlaps with general Bethe states that was obtained in the context of the XXZ chain with open boundary conditions \cite{Pozsgay_1309.4593, 2012_Kozlowski_JSTAT_P05021, 1998_Tsuchiya_JMathPhys_39}. The statement that overlaps with non-parity-invariant Bethe states vanish is still valid in the scaling limit to LL which means that the BEC state~\cite{LLpaper} has zero overlap with non-parity-invariant LL Bethe states. 
\end{abstract}


\section{Introduction}


The study of out-of-equilibrium dynamics of isolated many-body quantum systems has seen great progress during the last decade. In particular, much theoretical progress was made in the field of one-dimensional quantum integrable models. Out-of-equilibrium realizations of these models, despite being strongly-correlated \cite{GiamarchiBOOK} and having a complete set of algebraically independent conserved quantities, are believed to exhibit relaxation behaviour. 

This relaxation can be described, in principle, by a generalized Gibbs ensemble (GGE) \cite{2008_Rigol}. An alternative way was recently proposed in the context of a quantum quench, the so-called ``quench action approach'' \cite{2013_Caux_PRL_110}. This first-principles-based analysis of time-evolved expectation values of physical observables uses as input the overlaps of the initial state with energy eigenstates. After a saddle-point approximation that is exact in the thermodynamic limit, it predicts both the time evolution of expectation values of physical observables and their long-time stationary values. To obtain the latter, one in particular needs the leading order term of the overlaps in the thermodynamic limit.

Despite the fact that Slavnov's theorem \cite{1990_Slavnov_TMP_79_82} gives overlaps between eigenstates of a Bethe Ansatz-solvable \cite{1931_Bethe} Hamiltonian with (in principle) generic states, until recently very little was known about overlaps between eigenstates of two Hamiltonians of the same family but at different values of interaction (here, we primarily focus on two XXZ chains with different anisotropies, one of which is in the Ising limit). It is these overlaps that are required for the application of the quench action approach to interaction quenches. For the Lieb-Liniger model, exact overlaps of the ground state for an even number of free bosons (Bose-Einstein condensate (BEC) state) and energy eigenstates at generic finite interaction parameter were conjectured in \cite{LLpaper} and thereafter proven in \cite{XXZ_LL_proof}. Their form resembled the Gaudin determinant form for norms of Bethe states \cite{1981_Gaudin_PRD_23, 1982_Korepin}, and was suitable for analysis in the thermodynamic limit \cite{LLpaper}. This made an application of the quench action approach possible, leading to a description of the stationary state at late times after the quench. Surprisingly, and in contrast to thermal properties of the Lieb-Liniger gas, this description turned out to yield a closed-form solution in the thermodyamic limit, for any (repulsive) value of the final interaction strength. Of particular interest was the conclusion that these results remain inaccessible to GGE-based calculations \cite{2013_Kormos_PRB, LLpaper}. 

In the context of spin chains, in \cite{XXZpaper} overlaps between the N\'eel state and parity-invariant XXZ Bethe states with an even number of down spins were derived from an earlier form of the overlaps  \cite{1998_Tsuchiya_JMathPhys_39, 2012_Kozlowski_JSTAT_P05021, Pozsgay_1309.4593}. This new form has the same Gaudin-like structure as the Lieb-Liniger overlaps and therefore can be treated in the thermodynamic limit as done in \cite{XXZpaper, Neel_to_XXZ_quench_paper}. Note that this result is still restricted to an even number of down spins.

In this paper we present a derivation of the overlaps of the zero-momentum N\'eel state and parity-invariant Bethe states, but now in the sectors of odd numbers of down spins. We also take the scaling limit to the Lieb-Liniger model and present the overlaps between the BEC state and Bose gas Bethe states with an odd number of bosons. The calculations are not trivial and the results show a slightly different structure of the determinant as compared to the even particle case \cite{XXZpaper}. These new overlap formulas may be specifically useful in broader contexts, such as in their relation to exact solutions of the 1D Kardar-Parisi-Zhang equation \cite{Calabrese_1402.1278}.

Furthermore, unlike the Lieb-Liniger case, up to now a proof that non-parity-invariant XXZ Bethe states have vanishing overlaps with the initial N\'eel state was lacking. In this paper we present this proof, both for an even and odd number of down spins. The relevance for the quench action approach is clear, as the time evolution of expectation values is governed by a double sum over the complete Hilbert space, including non-parity-invariant states. 

The paper is organized as follows. In section \ref{sec:XXZchain} we review the Bethe Ansatz solution of the XXZ chain and recall the Gaudin norm formula of Bethe states \cite{1981_Gaudin_PRD_23}, which will be used in subsequent sections. In section \ref{sec:oddandnpi} we present the derivation of the overlap formula with parity-invariant Bethe states with an odd number of down spins, and we prove that non-parity-invariant states have zero overlap with the N\'eel state. In section \ref{sec:scaling_LL} we perform the scaling limit to the Lieb-Liniger Bose gas and present the BEC overlap formula for an odd number of bosons.

\section{The spin-1/2 XXZ chain} \label{sec:XXZchain}
The model which we initially focus on is the well-known spin-$1/2$ anisotropic Heisenberg chain with Hamiltonian
\begin{equation}\label{eq:Hamiltonian_XXZ}
	H = J\sum_{j=1}^{N}\left(\sigma_{j}^{x}\sigma_{j+1}^{x}+\sigma_{j}^{y}\sigma_{j+1}^{y}+\Delta ( \sigma_{j}^{z}\sigma_{j+1}^{z}-1)\right)\epp
\end{equation}
Since we are interested in antiferromagnetic cases, the overall exchange coupling constant is taken to be positive, $J>0$.
The exchange anisotropy is parametrized as $\Delta=\cosh(\eta)$ with $\eta \in \mathbb{C}$ being a free parameter (we will in practice of course limit ourselves to real $\Delta$).
We consider a chain with $N$ sites (which we choose to be an even number) and impose periodic boundary conditions $\sigma_{N+1}^\alpha=\sigma_1^\alpha$, $\alpha=x,y,z$.
 
This Hamiltonian can be diagonalized by Bethe Ansatz \cite{1931_Bethe, 1958_Orbach_PR_112}. Starting from the ferromagnetic reference state $\left|\uparrow\right\rangle^{\otimes N}=\left|\uparrow \uparrow \ldots \uparrow \right\rangle$, we can construct eigenstates with $M\leq N/2$ overturned spins as
\begin{subequations}\label{eq:BA_state}
\begin{equation}
	|\{\lambda_j\}_{j=1}^M\rangle = \sum\nolimits_{\{s_j\}_{j=1}^M \subset \{1,\ldots,N\}} \Psi_{M}\!\left(\{s_j\}_{j=1}^M|\{\lambda_j\}_{j=1}^M\right)\ \sigma_{s_1}^-\ldots\sigma_{s_M}^-|\uparrow\uparrow\ldots\uparrow\rangle
\end{equation}
where the positions of the downturned spins are labeled by the indices $s_j$, $j=1,\ldots,M$, in such a way that $s_j<s_k$ for $j<k$. The amplitudes take the Bethe Ansatz form
\begin{equation} \label{eq:BA_state_wf}
\Psi_{M}\!\left(\{s_j\}_{j=1}^M|\{\lambda_j\}_{j=1}^M\right)=\sum_{Q\in\mathcal{S}_M} (-1)^{[Q]} 
 \exp\left\{- i \sum_{a=1}^M p(\lambda_{Q_a})s_a - \frac{i}{2} \sum_{\substack{ a,b=1\\ b>a}}^M \theta(\lambda_{Q_b} - \lambda_{Q_a}) \right\}\epp
\end{equation}
\end{subequations}
Here, $\mathcal{S}_M$ represents the set of all permutations of $1, \ldots, M$, and $(-1)^{[Q]}$ represents the parity of the permutation. The state \eqref{eq:BA_state} is an eigenstate of the total momentum operator with eigenvalue
\begin{subequations}
\begin{equation}\label{eq:total_momentum}
P = \sum_{j=1}^M p(\lambda_j)\epc \quad\text{where}\quad p(\lambda)= - i\ln \left[\frac{\sinh(\lambda + \eta/2)}{\sinh(\lambda- \eta/2)}\right]\epp
\end{equation}
The scattering phase shift in the wave function \eqref{eq:BA_state_wf} is given by
\begin{equation}
	\theta(\lambda) = i\ln\left(\frac{\sinh(\lambda+\eta)}{\sinh(\lambda-\eta)}\right)\epp
\end{equation}
\end{subequations}
The branches of the logarithms are fixed by $p(0^\pm)=\pm\pi$ and $\theta(0^\pm)=\mp\pi$. The quasi-momenta $\{\lambda_j\}_{j=1}^M$, which form a self-conjugate set \cite{1986_Vladimirov_TMP_66}, are called rapidities or Bethe roots, and they specify an eigenstate of Hamiltonian \eqref{eq:Hamiltonian_XXZ} with energy eigenvalue
\begin{equation}
E = \sum_{j=1}^M \frac{2J \sinh^2(\eta)}{\sinh(\lambda_j + \eta/2) \sinh(\lambda_j - \eta/2) } \epc
\end{equation}
under the condition that they fulfill the Bethe equations
\begin{equation}\label{eq:BAE}
	\left(\frac{\sinh(\lambda_j+\eta/2)}{\sinh(\lambda_j-\eta/2)}\right)^N=-\prod_{k=1}^M\frac{\sinh(\lambda_j-\lambda_k+\eta)}{\sinh(\lambda_j-\lambda_k-\eta)}\epc \qquad j=1,\ldots,M \epp
\end{equation}
The norm of an eigenstate is given in terms of the Gaudin determinant \cite{1981_Gaudin_PRD_23,1982_Korepin}
\begin{subequations}\label{eq:norm_Bethe_state}
\begin{align}
\|\{\lambda_j\}_{j=1}^M\| &= \sqrt{\langle \{\lambda_j\}_{j=1}^M|  \{\lambda_j\}_{j=1}^M \rangle}\epc \\ 
\label{eq:norm_Bethe_state_b}
	\langle \{\lambda_j\}_{j=1}^M|  \{\lambda_j\}_{j=1}^M \rangle &= \sinh^M(\eta) \prod_{\substack{j,k=1\\j\neq k}}^M \frac{\sinh(\lambda_j - \lambda_k + \eta)}{\sinh(\lambda_j - \lambda_k)} \det{}_{\!M} (G_{jk}) \epc\\
	G_{jk} &= \label{eq:Gaudin_matrix}  \delta_{jk}\left(NK_{\eta/2}(\lambda_j)-\sum_{l=1}^{M}K_\eta(\lambda_j-\lambda_l)\right) + K_\eta(\lambda_j-\lambda_k)\epc
\end{align}
\end{subequations}
where $K_\eta(\lambda)=\frac{\sinh(2\eta)}{\sinh(\lambda+\eta)\sinh(\lambda-\eta)} = i\partial_{\lambda} \theta(\lambda)$.

We shall write that a Bethe state of the form \eqref{eq:BA_state} is ``on-shell'' if the parameters $\{\lambda_j\}_{j=1}^M$ satisfy Bethe equations \eqref{eq:BAE}. In contrast, we will write ``off-shell'' if they are arbitrary complex numbers.  We further call a Bethe state parity-invariant if they obey the symmetry condition $\{\lambda_j\}_{j=1}^M= \{-\lambda_j\}_{j=1}^M$, where, for generic $\Delta=\cosh(\eta)$, i.e.~$\Delta>1$ or $-1 \leq \Delta =(q+q^{-1})/2\leq 1$, $q=e^\eta$ not a root of unity, we identify rapidities when their imaginary parts differ by $\pm\pi$. This in particular means that the two points $\pm i\pi/2$ are identified. Hence, adding a single rapidity at $i\pi/2$ to a parity-invariant set of rapidities does not destroy this symmetry. In the root-of-unity case, e.g.~when $\Delta\to\pm 1$, we consider rescaled rapidities, e.g.~$\tilde{\lambda_j} = \lambda_j/(i\eta)$ or $\tilde{\lambda_j} = \lambda_j/(i\pi-i\eta)$ respectively, and we identify
the two points $\pm\infty$, because in these special cases Bethe equations allow for these type of solutions.

\section{Overlap of the N\'eel state with zero-magnetization XXZ Bethe states} \label{sec:oddandnpi}
In the following we are interested in overlaps of zero-magnetization XXZ Bethe states with the zero-momentum N\'eel state, which is the symmetric combination of N\'eel and anti-N\'eel. Whenever we mention in the following the N\'eel state $|\Psi_0\rangle$ we mean this linear combination with momentum zero. 

Before proceeding further, let us define four classes of zero-momentum Bethe states, and mention where they are specifically treated:
\begin{itemize}
	\item Parity-invariant Bethe states with an even number of down spins (Sec \ref{sec:pi_even}, Ref.~\cite{XXZpaper}),
	\item Parity-invariant Bethe states with an odd number of down spins (Sec.~\ref{sec:pi_odd}),
	\item Non-parity-invariant Bethe states without any Bethe root at zero (Sec.~\ref{sec:npi_even}),
	\item Non-parity-invariant Bethe states with one Bethe root at zero (Sec.~\ref{sec:npi_odd}).
\end{itemize}
The number of downturned spins is given by the number of Bethe roots $M$. In this section we only consider Bethe states lying in the zero-magnetization sector ($M=N/2$). The nonzero magnetization sector shall become important for the scaling to the LL model in Sec.~\ref{sec:scaling_LL}.

\subsection{Parity-invariant Bethe states}\label{sec:pi}
We denote a parity-invariant (off-shell) Bethe state with an even number of down spins by $|\{\pm\mu_j\}_{j=1}^{M/2}\rangle$. In Ref.~\cite{XXZpaper}, an explicit formula for the overlap $\langle \Psi_0 | \{\pm\mu_j\}_{j=1}^{M/2}\rangle$ of the N\'eel state 
\begin{equation}
	|\Psi_0\rangle = \frac{1}{\sqrt{2}}\left(\left|\uparrow\downarrow \right\rangle^{\otimes N/2} + \left|\downarrow\uparrow \right\rangle^{\otimes N/2} \right)
\end{equation}
with parity-invariant Bethe states in the zero-magnetization sector $M=N/2$ was given (See Sec.~\ref{sec:pi_even}). This could be extended to the sector of nonzero magnetization ($M<N/2$) in Ref.~\cite{XXZ_LL_proof}, where the overlaps with so-called $q$-raised N\'eel states were derived. However, these results are still restricted to even numbers of down spins. 

In Sec.~\ref{sec:pi_odd} we present an explicit expression for the overlap of the N\'eel state with a Bethe state with an odd number of parity-invariant Bethe roots valid for generic $\Delta$. The result shows a slightly different structure of the determinant as compared to the case of even number of down spins. Note that an odd number of down spins implies that there is one Bethe root at the origin or at $i\pi/2$, the rest being arranged in symmetric pairs (around the origin). Thereafter we shall present a proof that non-parity-invariant Bethe states have no overlap with the N\'eel state for both cases, even and odd numbers of Bethe roots. 

\subsubsection{Even number of down spins.}\label{sec:pi_even}
The overlap of the zero-momentum N\'eel state with an unnormalized parity-invariant XXZ off-shell state with an even number of down spins is given by \cite{XXZpaper} ($N$ divisible by four, number of particles $M=N/2$ even)
\begin{subequations}\label{eq:overlap_XXZ_offshell} 
\begin{equation}
	 \langle \Psi_0 |\{\pm\lambda_j\}_{j=1}^{N/4}\rangle =  \gamma\det{}_{\!N/4}\left(G_{jk}^{+}\right)\epc
\end{equation}
where the prefactor $\gamma$  and the matrix $G_{jk}^+$ read
\begin{align}\label{eq:gamma}
\gamma &= \sqrt{2}\left[\prod_{j=1}^{N/4}\frac{\sine{\lambda_j}{\eta/2}\sine{\lambda_j}{-\eta/2}}{\sine{2\lambda_j}{0}^2}\right]\left[\prod_{\substack{j>k=1\\ \ \sigma=\pm}}^{N/4}\frac{\sine{\lambda_j+\sigma\lambda_k}{\eta}\sine{\lambda_j+\sigma\lambda_k}{-\eta}}{\sine{\lambda_j+\sigma\lambda_k}{0}^2}\right]\epc \\[3ex] 
G_{jk}^{+} &= \delta_{jk}\left(N\sine{0}{\eta}K_{\eta/2}(\lambda_j)-\sum_{l=1}^{N/4}\sine{0}{\eta}K_\eta^{+}(\lambda_j,\lambda_l)\right) + \sine{0}{\eta}K_\eta^{+}(\lambda_j,\lambda_k)\notag \\ \label{eq:Gaudin_plus}
&\qquad\quad + \delta_{jk}\frac{\sine{2\lambda_j}{\eta}\,\mathfrak{A}_j+\sine{2\lambda_j}{-\eta}\,\bar{\mathfrak{A}}_j}{\sine{2\lambda_j}{0}} + (1-\delta_{jk})f_{jk}\epc \quad\qquad j,k=1,\ldots,N/4\epc\\[3ex] \label{eq:f_jk}
f_{jk} 
&= \mathfrak{A}_k\left( \frac{\sine{2\lambda_j}{\eta} \sine{0}{\eta}}{\sine{\lambda_j+\lambda_k}{0}\sine{\lambda_j-\lambda_k}{\eta}} - \frac{\sine{2\lambda_j}{-\eta}\sine{0}{\eta}}{\sine{\lambda_j-\lambda_k}{0}\sine{\lambda_j+\lambda_k}{-\eta}} \right) + \mathfrak{A}_k\bar{\mathfrak{A}}_j \left(\frac{\sine{2\lambda_j}{-\eta}\sine{0}{\eta}}{\sine{\lambda_j-\lambda_k}{0}\sine{\lambda_j+\lambda_k}{-\eta}}\right) \notag\\
&\quad - \bar{\mathfrak{A}}_j\left(\frac{\sine{2\lambda_j}{-\eta}\sine{0}{\eta}}{\sine{\lambda_j-\lambda_k}{0}\sine{\lambda_j+\lambda_k}{-\eta}} + \frac{\sine{2\lambda_j}{-\eta}\sine{0}{\eta}}{\sine{\lambda_j+\lambda_k}{0}\sine{\lambda_j-\lambda_k}{-\eta}}\right)
\end{align}
\end{subequations}
with 
\begin{equation}
 K_\eta^{+}(\lambda,\mu) = K_\eta(\lambda+\mu)+K_\eta(\lambda-\mu)
 \quad\text{and}\quad K_\eta(\lambda)=\frac{\sine{0}{2\eta}}{\sine{\lambda}{\eta}\sine{\lambda}{-\eta}}\epp
\end{equation}
We also introduced the abbreviations 
\begin{equation}\label{eq:def_sine}
\sine{\lambda}{\eta}=\sinh(\lambda+\eta)
\end{equation}
and
\begin{equation}\label{eq:func_a_tilde}
	\mathfrak{A}_j = 1 + \mathfrak{a}_j\epc\quad \bar{\mathfrak{A}}_j = 1 + \mathfrak{a}_j^{-1}\epc\quad \mathfrak{a}_j = \left[\prod_{\substack{k=1\\ \ \sigma=\pm}}^{N/4}\frac{\sine{\lambda_j-\sigma\lambda_k}{-\eta}}{\sine{\lambda_j-\sigma\lambda_k}{\eta}}\right]\left(\frac{\sine{\lambda_j}{\eta/2}}{\sine{\lambda_j}{-\eta/2}}\right)^{N}\epp
\end{equation}
Formula \eqref{eq:overlap_XXZ_offshell} still holds for the bra states $\langle N | = \left\langle\uparrow\downarrow \right|^{\otimes N/2}$ or $\langle AN | = \left\langle\downarrow\uparrow\right|^{\otimes N/2}$. Note that then $\gamma$ has to be modified by a factor $1/\sqrt{2}$. 

\subsubsection{Odd number of down spins.}\label{sec:pi_odd}
We consider the case when an odd number of spins are turned down. We choose $N-2$ divisible by four, $M = N/2$ odd. A parity-invariant off-shell Bethe state has either one Bethe root at the origin or at $i\pi/2$. The former leads to a state with total momentum $P=\pi$, the latter to a state with $P=0$ [see Eq.~\eqref{eq:total_momentum}]. Since we are interested in the overlap with the zero-momentum N\'eel state $|\Psi_0\rangle$, only the latter matters. We define $M' = (M-1)/2$ as the number of Bethe roots with positive real part. The overlap of the N\'eel state $|\Psi_0\rangle$ with these (unnormalized) parity-invariant off-shell states is (see \ref{sec:odd_proof}, $\lambda_0 = 0, i\pi/2$)
\begin{subequations}\label{eq:overlap_XXZ_offshell_odd} 
\begin{equation}
	 \langle \Psi_0 |\{\pm\lambda_j\}_{j=1}^{M'} \cup\{\lambda_0\}\rangle =  \gamma_{odd}\det{}_{\!M'+1}\left(G_{jk}^{(+,odd)}\right)\epc
\end{equation}
where the prefactor $\gamma_{odd}$ and the matrix $G_{jk}^{(+,odd)}$ read
\begin{align}\label{eq:gamma_odd}
\gamma_{odd} &= \sqrt{2}\frac{\sinh(\frac{\eta}{2})}{\sinh(\eta)}\left[\prod_{j=1}^{M'}\frac{\sine{\lambda_j}{\eta/2}\sine{\lambda_j}{-\eta/2}}{\sine{2\lambda_j}{0}^2}\frac{\sine{\lambda_j}{\eta}\sine{\lambda_j}{-\eta}}{\sine{\lambda_j}{0}^2}\right]\left[\prod_{\substack{j>k=1\\ \ \sigma=\pm}}^{M'}\frac{\sine{\lambda_j+\sigma\lambda_k}{\eta}\sine{\lambda_j+\sigma\lambda_k}{-\eta}}{\sine{\lambda_j+\sigma\lambda_k}{0}^2}\right]\epc\\[3ex] 
G_{jk}^{(+,odd)} &= \delta_{jk}\left(N\sine{0}{\eta}K(j)-\sum_{l=0}^{M'}\sine{0}{\eta}K(j,l)\right) + \sine{0}{\eta}K(j,k)\notag \\ \label{eq:Gaudin_plus_odd}
&\quad + \delta_{jk}(1-\delta_{j0})\frac{\sine{2\lambda_j}{\eta}\,\mathfrak{A}_j+\sine{2\lambda_j}{-\eta}\,\bar{\mathfrak{A}}_j}{\sine{2\lambda_j}{0}} + (1-\delta_{jk})f_{jk}\epc \quad j,k=0,\ldots,M'\epc\\[3ex] \label{eq:f_jk_odd}
f_{jk} 
&= \mathfrak{A}_k\left( \frac{\sine{2\lambda_j}{\eta} \sine{0}{\eta}}{\sine{\lambda_j+\lambda_k}{0}\sine{\lambda_j-\lambda_k}{\eta}} - \frac{(1-\delta_{j0})\sine{2\lambda_j}{-\eta}\sine{0}{\eta}}{\sine{\lambda_j-\lambda_k}{0}\sine{\lambda_j+\lambda_k}{-\eta}} \right) + \mathfrak{A}_k\bar{\mathfrak{A}}_j \left(\frac{\sine{2\lambda_j}{-\eta}\sine{0}{\eta}}{\sine{\lambda_j-\lambda_k}{0}\sine{\lambda_j+\lambda_k}{-\eta}}\right) \notag\\
&\quad - \bar{\mathfrak{A}}_j\left(\frac{\sine{2\lambda_j}{-\eta}\sine{0}{\eta}}{\sine{\lambda_j-\lambda_k}{0}\sine{\lambda_j+\lambda_k}{-\eta}} + \frac{(1-\delta_{0k})\sine{2\lambda_j}{-\eta}\sine{0}{\eta}}{\sine{\lambda_j+\lambda_k}{0}\sine{\lambda_j-\lambda_k}{-\eta}}\right)\epc \\[2ex]
K(j,k) &= 
\left\{ \begin{array}{r@{\hspace{4ex}}l} \left(K_{\eta}(\lambda_j-\lambda_k) + K_{\eta}(\lambda_j+\lambda_k)\right)\epc & 1\leq j,k \leq M'\epc \\ 
\frac{1}{2}
\left(K_{\eta}(\lambda_j-\lambda_k) + K_{\eta}(\lambda_j+\lambda_k)\right) \epc & \text{otherwise}\epc
\end{array}   \right.
\end{align}
and $K(j)=\frac{1}{1+\delta_{j0}} K_{\eta/2}(\lambda_j)$. 
The abbreviations $\mathfrak{A}_k$ and $\bar{\mathfrak{A}}_j$ now read
\begin{equation}\label{eq:func_a_tilde_odd}
	\mathfrak{A}_k = 1 + \mathfrak{a}_k\epc\quad \bar{\mathfrak{A}}_j = 1 + \mathfrak{a}_j^{-1}\epc\quad \mathfrak{a}_j = \frac{\sine{\lambda_j}{-\eta}}{\sine{\lambda_j}{\eta}}\left[\prod_{\substack{k=1\\ \ \sigma=\pm}}^{M'}\frac{\sine{\lambda_j-\sigma\lambda_k}{-\eta}}{\sine{\lambda_j-\sigma\lambda_k}{\eta}}\right]\left(\frac{\sine{\lambda_j}{\eta/2}}{\sine{\lambda_j}{-\eta/2}}\right)^{N}\epp
\end{equation}
\end{subequations}  
Note the additional factors $(1-\delta_{j0})$ in the ``on-shell-vanishing'' terms and the factors $1/2$ in the functions $K(j,k)$ and $K(j)$ as soon as one of the arguments $j,k$ is $0$ which addresses the parameter $\lambda_0=0,i\pi/2$. 

For on-shell Bethe states all $\mathfrak{A}_k$, $\bar{\mathfrak{A}}_j$ terms vanish. Using norm formula \eqref{eq:norm_Bethe_state} we can write the overlap of the N\'eel state $|\Psi_0\rangle$ with normalized on-shell states as
\begin{subequations}\label{eq:overlap_odd}
\begin{align}\label{eq:overlap_a}
	 \frac{\langle \Psi_0|\{\pm\lambda_j\}_{j=1}^{M'}\cup \{\lambda_0=i\pi/2\} \rangle}{ \|\{\pm\lambda_j\}_{j=1}^{M'} \cup \{i\pi/2\} \|} &= \sqrt{2}\frac{\sinh(\frac{\eta}{2})}{\sinh(\eta)}\left[\prod_{j=1}^{M'}\frac{\sqrt{\tanh(\lambda_j+\frac{\eta}{2}) \tanh(\lambda_j-\frac{\eta}{2})}}{2\sinh(2\lambda_j)}\right]
\notag\\ 
&\quad\times\frac{\det{}_{\!M'+1}(G_{jk}^{+})}{\sqrt{\det{}_{\!2M'+1}(G_{jk})}}\epc
\end{align}
where
\begin{align}
G_{jk}^+ &= \delta_{jk}\left(N\sine{0}{\eta}K(j)-\sum_{l=0}^{M'}\sine{0}{\eta}K^+(j,l)\right) + \sine{0}{\eta}K^+(j,k)\epc \quad 0\leq j,k\leq M'\epc \\ 
G_{jk} &= \delta_{jk}\left(N\sine{0}{\eta}K_{\eta/2}(\lambda_j)-\sum_{l=0}^{2M'}\sine{0}{\eta}K_\eta(\lambda_j-\lambda_l)\right) + \sine{0}{\eta}K_\eta(\lambda_j-\lambda_k)\epc\notag\\ &\hspace{54.4ex} \quad 0\leq j,k\leq 2M'\epp
\end{align}
\end{subequations}
This completes the previous result (see Sec.~\ref{sec:pi_even} and Ref.~\cite{XXZpaper}) of an even number of downturned spins.

\subsection{Non-parity-invariant Bethe states}\label{sec:npi}
In this subsection we prove that non-parity-invariant on-shell Bethe states have no overlap with the N\'eel state $|\Psi_0\rangle$ in both cases, even and odd numbers of down spins. In Refs.~\cite{Pozsgay_1309.4593, 2012_Kozlowski_JSTAT_P05021} it is shown that overlaps of the states $|N\rangle =\left|\uparrow\downarrow\right\rangle^{\otimes M}$ or $|AN\rangle=\left|\downarrow\uparrow\right\rangle^{\otimes M}$ with an off-shell Bethe state $|\lambda\rangle =|\{\lambda_j\}_{j=1}^M\rangle$ are given by the following determinant expression (note that we here use a different normalization of Bethe states as compared to Ref.~\cite{Pozsgay_1309.4593}), 
\begin{subequations}\label{eq:overlap_S3}
\begin{align}\label{eq:overlap_S3a}
	e^{iP}\langle AN|\lambda\rangle &= \langle N|\lambda\rangle = \left[\prod_{j=1}^M\frac{\sine{\lambda_j}{\eta/2}}{\sine{2\lambda_j}{0}}\;\frac{\sine{\lambda_j}{-\eta/2}^{M}}{ \sine{\lambda_j}{\eta/2}^{M}} \right] \left[\prod_{j>k=1}^{M}\frac{\sine{\lambda_j+\lambda_k}{\eta}}{\sine{\lambda_j+\lambda_k}{0}}\right]\det{}_{\!M}(\mathds{1}+U)\epc\\
	\label{eq:overlap_S3b}
	U_{jk} &= \frac{\sine{2\lambda_k}{\eta}\sine{2\lambda_k}{0}}{\sine{\lambda_j+\lambda_k}{0}\sine{\lambda_j-\lambda_k}{\eta}}\left[\prod_{\substack{l=1\\ l\neq k}}^{M}\frac{\sine{\lambda_k+\lambda_l}{0}}{\sine{\lambda_k-\lambda_l}{0}}\right]\left[\prod_{l=1}^M\frac{\sine{\lambda_k-\lambda_l}{-\eta}}{\sine{\lambda_k+\lambda_l}{\eta}}\right]\left(\frac{\sine{\lambda_k}{\eta/2}}{\sine{\lambda_k}{-\eta/2}}\right)^{2M}\epc
\end{align}
\end{subequations}
where $\sine{\lambda}{\eta}$ is defined in \eqref{eq:def_sine} and $P=0,\pi$ is the total momentum \eqref{eq:total_momentum} of the Bethe state. Note that there is a difference of a factor $\sqrt{2}$ compared to the usual expression because here we consider overlaps with N\'eel and anti-N\'eel states instead of its symmetric combination. The parameters $\lambda_j$, $j=1,\ldots, M$, are arbitrary complex numbers. We show that the determinant of $\mathds{1}+U$ is zero for non-parity-invariant on-shell Bethe states, {\it i.e.} the set $\{\lambda_j\}_{j=1}^M$ are Bethe roots and there is at least one $j$ for which $\lambda_j \neq -\lambda_k$ for all $k=1,\ldots,M$. We multiply $\mathds{1}+U$ from the right with the diagonal matrix 
\begin{equation}
	D = \text{diag}\left\{ \left[ \prod\limits_{\substack{l=1\\ l\neq k}}^M \frac{\sine{\lambda_k + \lambda_l}{\eta}}{\sine{\lambda_k - \lambda_l}{ - \eta}} \right]\left(\frac{\sine{\lambda_k}{-\eta/2}}{\sine{\lambda_k}{\eta/2}}\right)^{2M}\right\}_{k=1}^M\epp
\end{equation}
The resulting matrix $B=(B_{jk})$ reads
\begin{align}\label{eq:B_matrix}
	B_{jk} &= \delta_{jk} \left[\prod_{\substack{l=1\\ l\neq j}}^{M}\frac{\sine{\lambda_j+\lambda_l}{\eta}}{\sine{\lambda_j-\lambda_l}{\eta}}\right] \left[-\prod_{\substack{l=1}}^{M}\frac{\sine{\lambda_j-\lambda_l}{\eta}}{\sine{\lambda_j-\lambda_l}{-\eta}}\right] \left(\frac{\sine{\lambda_j}{-\eta/2}}{\sine{\lambda_j}{\eta/2}}\right)^{2M}- \frac{\sine{0}{\eta}\sine{2\lambda_k}{0}}{\sine{\lambda_j+\lambda_k}{0}\sine{\lambda_j-\lambda_k}{\eta}}\prod_{\substack{l=1\\ l\neq k}}^M\frac{\sine{\lambda_k+\lambda_l}{0}}{\sine{\lambda_k-\lambda_l}{0}}\notag\\ 
	&= \delta_{jk} b_j\left[\prod_{\substack{l=1\\ l\neq j}}^{M}\frac{\sine{\lambda_j+\lambda_l}{\eta}}{\sine{\lambda_j-\lambda_l}{\eta}}\right] - \frac{\sine{0}{\eta}\sine{2\lambda_k}{0}}{\sine{\lambda_j+\lambda_k}{0}\sine{\lambda_j-\lambda_k}{\eta}}\prod_{\substack{l=1\\ l\neq k}}^M\frac{\sine{\lambda_k+\lambda_l}{0}}{\sine{\lambda_k-\lambda_l}{0}}\epc
\end{align}
where we defined the abbreviations
\begin{equation}\label{eq:def_bj}
	b_j = b(\lambda_j)\epc\ j=1,\ldots,M\epc \qquad \text{with}\qquad b(\lambda) = 	\left[-\prod_{\substack{l=1}}^{M}\frac{\sine{\lambda-\lambda_l}{\eta}}{\sine{\lambda-\lambda_l}{-\eta}}\right] \left(\frac{\sine{\lambda}{-\eta/2}}{\sine{\lambda}{\eta/2}}\right)^{2M}\epp
\end{equation}
The overlap prefactor in $\langle N | \{\lambda_j\}_{j=1}^M\rangle = \gamma_B \det{}_{\!M}(B)$ reads now 
\begin{equation}
	\gamma_B = \left[\prod_{j=1}^M\frac{\sine{\lambda_j}{\eta/2}}{\sine{2\lambda_j}{0}}\;\frac{\sine{\lambda_j}{\eta/2}^{M}}{ \sine{\lambda_j}{-\eta/2}^{M}} \right] \left[\prod_{j>k=1}^{M}\frac{\sine{\lambda_j-\lambda_k}{\eta}\sine{\lambda_j-\lambda_k}{-\eta}}{\sine{\lambda_j+\lambda_k}{0}\sine{\lambda_j+\lambda_k}{\eta}}\right]\epp
\end{equation} 
It causes problems if one of the spectral parameters is at the origin (or at $i\pi/2$, or both) or if the set $\{\lambda_j\}_{j=1}^M$ contains one or more pairs of the form $\lambda_j=-\lambda_k$. The poles at $\lambda_j = \eta/2$, $\lambda_j = -\lambda_k -\eta$ are less important because at the end $\lambda_j$ are Bethe roots, and string solutions to the Bethe equations \eqref{eq:BAE} for finite $N$ show always deviations of perfect strings. Exceptions are strings with zero real part that belong to parity invariant states, but those are not considered here. In all cases of non-parity-invariant states we shall show that the product of $\det{}_{\!M}(B)$ and $\gamma_B$ vanishes if we impose the on-shell condition (Bethe equations \eqref{eq:BAE}). For this purpose we pull the divergent parts of $\gamma_B$ into the determinant and explicitly construct an eigenvector of the resulting matrix that has eigenvalue zero.

\subsubsection{States with only nonzero Bethe roots.} \label{sec:npi_even}
Let us first consider the case with $m$ pairs $\lambda_{2j-1}=-\lambda_{2j} =: \mu_j\neq 0,i\pi/2$, $j=1,\ldots, m$, and all other Bethe roots are unpaired: $\lambda_j \neq -\lambda_k,$ and $\lambda_j\neq 0,i\pi/2$ for all $j,k = 2m+1, \ldots, M$. The case of one Bethe root at zero or $i\pi/2$ is treated in the next subsection, Sec.~\ref{sec:npi_odd}. We start with the determinant expression for off-shell Bethe states using matrix $B$ defined in Eq.~\eqref{eq:B_matrix}. We impose Bethe equations later. We redefine $\lambda_{2j-1} = \mu_j + \epsilon_j$ and $\lambda_{2j} = -\mu_j + \epsilon_j$ for $j=1,\ldots,m$ and consider the limits $\epsilon_j\to 0$, $j=1,\ldots,m$, when the subset $\{\lambda_j\}_{j=1}^{2m}$ of $\lambda$'s becomes parity invariant. We have
\begin{align}\label{eq:PpPm}
	b_{2j-1}b_{2j} = b(\mu_j+\epsilon_j)b(-\mu_j+\epsilon_j) &= \left[\prod_{\substack{l=1\\j\neq 2j-1,2j}}^M\frac{\sine{\mu_j-\lambda_l}{\eta}}{\sine{\mu_j+\lambda_l}{\eta}}\right]\left[\prod_{\substack{l=1\\j\neq 2j-1,2j}}^M\frac{\sine{\mu_j+\lambda_l}{-\eta}}{\sine{\mu_j-\lambda_l}{-\eta}}\right] + \mathcal{O}(\epsilon_j)\notag \\ &= P_{+}^{-1}(\mu_j)P_{-}(\mu_j) + \mathcal{O}(\{\epsilon_j\}_{j=1}^m)\epc
\end{align}
where $P_{\sigma}(\lambda) = \prod_{k=2m+1}^M\frac{\sine{\lambda+\lambda_k}{\sigma\eta}}{\sine{\lambda-\lambda_k}{\sigma\eta}}$ for $\sigma = \pm, 0$. 
We write down the zeroth order in all $\{\epsilon_j\}_{j=1}^m \equiv \epsilon$ of the matrix elements $B_{jk}$, where ``order $\epsilon$'' or $\mathcal{O}(\epsilon)$ means $\mathcal{O}(\epsilon_1,\ldots,\epsilon_m)$, 
\begin{subequations}
\begin{align}
	B_{2j-1,2j-1} &= \frac{\sine{0}{\eta}}{\sine{2\mu_j}{\eta}} b(\mu_j) P_+(\mu_j) + \mathcal{O}(\epsilon)\epc \\
	B_{2j-1,2j\phantom{-1}} &= -\frac{\sine{0}{\eta}}{\sine{2\mu_j}{\eta}} P_0^{-1}(\mu_j) + \mathcal{O}(\epsilon)\epc\\
	B_{2j,2j-1\phantom{-1}} &= \frac{\sine{0}{\eta}}{\sine{2\mu_j}{-\eta}} P_0(\mu_j) + \mathcal{O}(\epsilon) \epc \\
	B_{2j,2j\phantom{-1}\phantom{-1}} &= -\frac{\sine{0}{\eta}}{\sine{2\mu_j}{-\eta}} b(-\mu_j) P_{-}^{-1}(\mu_j)  + \mathcal{O}(\epsilon) = -\frac{\sine{0}{\eta}}{\sine{2\mu_j}{-\eta}} b^{-1}(\mu_j) P_{+}^{-1}(\mu_j)  + \mathcal{O}(\epsilon) \epc \\
	B_{k,2j-1} &= \mathcal{O}(\epsilon) \epc \\
	B_{k,2j\phantom{-1}} &= \mathcal{O}(\epsilon) \epc \\
	B_{2j-1,k} &= -\frac{\sine{0}{\eta}\sine{2\lambda_k}{0}}{\sine{\mu_j-\lambda_k}{\eta}\sine{\mu_j+\lambda_k}{0}} \prod_{\substack{l=2m+1\\l\neq k}}^M\frac{\sine{\lambda_k+\lambda_l}{0}}{\sine{\lambda_k-\lambda_l}{0}}  + \mathcal{O}(\epsilon)\epc \\
	B_{2j,k\phantom{-1}} &= -\frac{\sine{0}{\eta}\sine{2\lambda_k}{0}}{\sine{\mu_j+\lambda_k}{-\eta}\sine{\mu_j-\lambda_k}{0}} \prod_{\substack{l=2m+1\\l\neq k}}^M\frac{\sine{\lambda_k+\lambda_l}{0}}{\sine{\lambda_k-\lambda_l}{0}}  + \mathcal{O}(\epsilon) \epc \\
	B_{jk} &= \delta_{jk} b_j\left[\prod_{\substack{l=2m+1\\ l\neq j}}^{M}\frac{\sine{\lambda_j+\lambda_l}{\eta}}{\sine{\lambda_j-\lambda_l}{\eta}}\right] - \frac{\sine{0}{\eta}\sine{2\lambda_k}{0}}{\sine{\lambda_j+\lambda_k}{0}\sine{\lambda_j-\lambda_k}{\eta}}\prod_{\substack{l=2m+1\\ l\neq k}}^M\frac{\sine{\lambda_k+\lambda_l}{0}}{\sine{\lambda_k-\lambda_l}{0}} + \mathcal{O}(\epsilon)\epp
\end{align}
\end{subequations}
It is $1\leq j \leq m$ and $k\neq 2j-1,2j$ everywhere, except for the last equation where $2m+1\leq j,k \leq M$. The $j$-th $2\times 2$ diagonal block is [$d_\pm \equiv \sine{0}{\eta}/\sine{2\mu_j}{\pm\eta}$, $b \equiv b(\mu_j)$, $P_\sigma \equiv P_\sigma(\mu_j)$, all these symbols evaluated at $\mu_j$] up to order $\epsilon$
\begin{multline}\label{eq:2x2_decomposition}
	\begin{pmatrix} d_+bP_+ & - d_+P_0^{-1}\\ d_- P_0 & - d_-(bP_+)^{-1} \end{pmatrix} 	= \begin{pmatrix} d_+\left(bP_+/P_0\right)^{1/2} & d_+\left(bP_+/P_0\right)^{1/2}\\ d_-\left(bP_+/P_0\right)^{-1/2} & d_-\left(bP_+/P_0\right)^{-1/2}\end{pmatrix}
	\\ \times
	\begin{pmatrix} \left(bP_+P_0\right)^{1/2} & 0\\ 0 & - \left(b P_+P_0\right)^{-1/2} \end{pmatrix}+\mathcal{O}(\epsilon)\epp
\end{multline}
The other elements in the corresponding two columns are of order $\epsilon$. We insert 
\begin{equation}\label{eq:res_of_identity}
	\begin{pmatrix} 1 & 0 \\ 0 & 1 \end{pmatrix} = \begin{pmatrix} 1 & 1-\alpha_j \\ -1 & \alpha_j \end{pmatrix} \begin{pmatrix} \alpha_j & \alpha_j-1 \\ 1 & 1 \end{pmatrix} 
\end{equation}
between the two matrices on the right, which yields again up to order $\epsilon$
\begin{equation}\label{eq:new_2x2_decomposition}
	\begin{pmatrix} \mathcal{O}(\epsilon) & d_+\left(bP_+/P_0\right)^{1/2} \\ \mathcal{O}(\epsilon) & d_-\left(bP_+/P_0\right)^{-1/2} \end{pmatrix}
	\begin{pmatrix} \alpha_j\left(bP_+P_0\right)^{1/2} & (1-\alpha_j)\left(b P_+P_0\right)^{-1/2}\\ \left(bP_+P_0\right)^{1/2} & - \left(b P_+P_0\right)^{-1/2} \end{pmatrix}\epp
\end{equation}
These manipulations of the first $m$ diagonal $2\times 2$ blocks affect also the first $2m$ columns, whereas the lower right $(M-2m)\times (M-2m)$ block of the $M\times M$ matrix $B$ remains unchanged. The important point of these determinant transformations is that, under the determinant,  we can absorb the divergent prefactors $\sine{0}{\eta}/\sine{2\epsilon_j}{0}$, $j=1,\ldots,m$, into the columns $2j-1$, $j=1,\ldots,m$, and then send all $\epsilon_j$'s to zero afterwards. We write the corresponding elements in columns $2j-1$ as $\ast$ because in these limits they are finite, but unimportant numbers. The corresponding elements in columns $2j$, $j=1,\ldots,m$, are all of order $\epsilon$ and hence go to zero in these limits. The determinant then reads
\begin{align}\label{eq:block_matrix_even}
\det{}_{\!M}(\tilde{B}) &= \lim\nolimits_{\{\epsilon_j\to 0\}_{j=1}^m} \prod_{j=1}^m\frac{\sine{0}{\eta}}{\sine{2\epsilon_j}{0}}\det{}_{\!M}(B) \notag\\
	&= \det{}_{\!M}\begin{pmatrix} \bar{B}_1  & O & \ldots & O & \begin{array}{l} \sss B_{1,2m+1} \\[-1.2ex] \sss B_{2,2m+1}\end{array} & \begin{array}{l} \sss\ldots \\[-1.2ex]\sss \ldots \end{array} & \begin{array}{l} \sss B_{1,M} \\[-1.2ex] \sss B_{2,M}\end{array} \\
O & \bar{B}_2 & \ldots & O & \begin{array}{l} \sss B_{3,2m+1} \\[-1.2ex] \sss B_{4,2m+1}\end{array} & \begin{array}{l} \sss\ldots \\[-1.2ex]\sss \ldots \end{array} & \begin{array}{l} \sss B_{3,M} \\[-1.2ex] \sss B_{4,M}\end{array} \\
\vdots & & \ddots & \vdots & \vdots & & \vdots \\
O & O & \ldots & \bar{B}_m & \begin{array}{l} \sss B_{2m-1,2m+1} \\[-1.2ex] \sss B_{2m,2m+1}\end{array} & \begin{array}{l} \sss\ldots \\[-1.2ex]\sss \ldots \end{array} & \begin{array}{l} \sss B_{2m-1,M} \\[-1.2ex] \sss B_{2m,M}\end{array} \\
O & O & \ldots & O & \begin{array}{l} \sss B_{2m+1,2m+1} \\[-1.2ex] \sss B_{2m+2,2m+1}\end{array} & \begin{array}{l} \sss\ldots \\[-1.2ex]\sss \ldots \end{array} & \begin{array}{l} \sss B_{2m+1,M} \\[-1.2ex] \sss B_{2m+2,M}\end{array} \\
\vdots & & & \vdots & \vdots & \ddots & \vdots \\
O & O & \ldots & O & \begin{array}{l} \sss B_{M-1,2m+1} \\[-1.2ex] \sss B_{M,2m+1}\end{array} & \begin{array}{l} \sss\ldots \\[-1.2ex]\sss \ldots \end{array} & \begin{array}{l} \sss B_{M-1,M} \\[-1.2ex] \sss B_{M,M}\end{array} \\
\end{pmatrix}\epc
\end{align}
where the symbol $O$ stands for $\begin{pmatrix} \ast & 0 \\ \ast & 0\end{pmatrix}$ and the diagonal blocks are given by
\begin{equation}
	\bar{B}_j = \begin{pmatrix}
		\ast & d_+(bP_+/P_0)^{1/2} \\
		\ast & d_-(bP_+/P_0)^{-1/2} 
	\end{pmatrix}\begin{pmatrix}
		\alpha_j(bP_+P_0)^{1/2} & (1-\alpha_j)(bP_+P_0)^{-1/2}\\
		(bP_+P_0)^{1/2} & -(bP_+P_0)^{-1/2}
	\end{pmatrix}\epp
\end{equation}
This determinant relation is valid in the off-shell sector. The parameters $\alpha_j$, $j=1,\ldots,m$, can be chosen arbitrarily. Note that one has to evaluate the functions $d_\pm$, $b$, $P_+$, and $P_0$ at the corresponding $\mu_j$. 

A determinant is the product of all its eigenvalues. We multiply matrix $\tilde{B}$ with one of its eigenvectors $(v_1,\ldots,v_M)^t$. The main idea of the steps from Eq.~\eqref{eq:2x2_decomposition} to Eq.~\eqref{eq:new_2x2_decomposition} is to choose the parameters $\alpha_j$ such that $v_{2j-1}\alpha_j[b(\mu_j)P_+(\mu_j)P_0(\mu_j)]^{1/2} + v_{2j}(1-\alpha_j)[b(\mu_j)P_+(\mu_j)P_0(\mu_j)]^{-1/2} = 0$. They exist if and only if 
\begin{equation}
	\frac{v_{2j}}{v_{2j-1}} \neq b(\mu_j)P_+(\mu_j)P_0(\mu_j)\epp
\end{equation}
Furthermore, we impose Bethe equations, {\it i.e.}~$b(\mu_j) = 1$ for $j=1,\ldots,m$ and $b_j = 1$ for $j=2m+1,\ldots,M$. Note that the symbols $b_j$ are defined in Eq.~\eqref{eq:def_bj} and correspond to the Bethe equations of the non-parity-invariant parameters $\lambda_j$, $j=2m+1,\ldots,M$ whereas $b(\mu_j)=1$ are the Bethe equations for the parameters $\mu_j$, corresponding to the parity-invariant subset of Bethe roots $\{\pm\mu_j\}_{j=1}^{m}$.

If the conditions $P_+(\mu_j)P_0(\mu_j) \neq 1$ are fulfilled for all $j=1,\ldots,m$, we can now show that $\vec{v} = (1,\ldots,1)^t$ is an eigenvector with eigenvalue zero. Multiplication of $\tilde{B}$ with $\vec{v}$ and using $\alpha_j[P_+(\mu_j)P_0(\mu_j)]^{1/2} +(1-\alpha_j)[P_+(\mu_j)P_0(\mu_j)]^{-1/2}= 0$ as well as $P_+(\mu_j)=P_-(\mu_j)$ [see Eq.~\eqref{eq:PpPm}] yields $(\tilde{B}\vec{v})_{2j-1} = f(\mu_j)$ and $(\tilde{B}\vec{v})_{2j} = g(\mu_j)$. The functions $f$ and $g$ are given by
\begin{subequations}
\begin{align}
	f(\mu) &= \frac{\sine{0}{\eta}}{\sine{2\mu}{\eta}}\prod_{l=2m+1}^M\frac{\sine{\mu+\lambda_l}{\eta}}{\sine{\mu-\lambda_l}{\eta}} - \frac{\sine{0}{\eta}}{\sine{2\mu}{\eta}}\prod_{l=2m+1}^M\frac{\sine{\mu-\lambda_l}{0}}{\sine{\mu+\lambda_l}{0}} \notag\\ 
	&\hspace{32ex} - \sum_{k=2m+1}^M\frac{\sine{0}{\eta}\sine{2\lambda_k}{0}}{\sine{\mu-\lambda_k}{\eta}\sine{\mu+\lambda_k}{0}}\prod_{\substack{l=2m+1\\l\neq k}}^M\frac{\sine{\lambda_k+\lambda_l}{0}}{\sine{\lambda_k-\lambda_l}{0}}\epc \\
	g(\mu) &= \frac{\sine{0}{\eta}}{\sine{2\mu}{-\eta}}\prod_{l=2m+1}^M\frac{\sine{\mu+\lambda_l}{0}}{\sine{\mu-\lambda_l}{0}} - \frac{\sine{0}{\eta}}{\sine{2\mu}{-\eta}}\prod_{l=2m+1}^M\frac{\sine{\mu-\lambda_l}{-\eta}}{\sine{\mu+\lambda_l}{-\eta}} \notag\\ 
	&\hspace{32ex} - \sum_{k=2m+1}^M\frac{\sine{0}{\eta}\sine{2\lambda_k}{0}}{\sine{\mu+\lambda_k}{-\eta}\sine{\mu-\lambda_k}{0}}\prod_{\substack{l=2m+1\\l\neq k}}^M\frac{\sine{\lambda_k+\lambda_l}{0}}{\sine{\lambda_k-\lambda_l}{0}}\epp \label{eq:g_function}
\end{align}
\end{subequations}
We show that they are identically zero for any set of complex parameters $\{\lambda_j\}_{j=2m+1}^M$. They are holomorphic in $\mathbb{C}$ because all residues vanish, 
\begin{subequations}
\begin{align}
0 &= \text{Res}(f,-\eta/2) = \text{Res}(f,-\lambda_k) =\text{Res}(f,\lambda_k-\eta) \epc \\
	0 &= \text{Res}(g,\eta/2) = \text{Res}(g,\lambda_k) =\text{Res}(g,-\lambda_k+\eta)
\end{align}
\end{subequations}
for all $k=2m+1,\ldots,M$. They are bounded and hence, with Liouville's theorem, we conclude that they are constant. Furthermore, $\lim_{\mu\to+\infty}f(\mu)=\lim_{\mu\to+\infty}g(\mu)=0$ and therefore $f(\mu)=0$ and $g(\mu)=0$ for all $\mu\in\mathbb{C}$. The components $(\tilde{B}\vec{v})_j$, $j=2m+1,\ldots,M$, can be written as $(\tilde{B}\vec{v})_j = h_j(\lambda_j)$ with
\begin{equation}
	h_j(\lambda) =\prod_{\substack{l=2m+1\\ l\neq j}}^M\frac{\sine{\lambda+\lambda_l}{\eta}}{\sine{\lambda-\lambda_l}{\eta}} -\prod_{\substack{l=2m+1\\l\neq j}}^M\frac{\sine{\lambda+\lambda_l}{0}}{\sine{\lambda-\lambda_l}{0}} + \sum_{\substack{k=2m+1\\ k\neq j}}^M\frac{\sine{0}{\eta}\sine{2\lambda_k}{0}}{\sine{\lambda-\lambda_k}{0}\sine{\lambda-\lambda_k}{\eta}}\prod_{\substack{l=2m+1\\l\neq k}}^M\frac{\sine{\lambda_k+\lambda_l}{0}}{\sine{\lambda_k-\lambda_l}{0}}\epp 
\end{equation}
The same reasoning as for $f,g$ applies for $h_j$. We also have vanishing residues, $\lim_{\lambda\to\infty}h_j(\lambda) = 0$, and therefore $h_j(\lambda)=0$ for all $\lambda\in \mathbb{C}$. 

In summary, we obtain $\tilde{B}\vec{v}= 0$. Therefore, $\det{}_{\!M}(\tilde{B})=0$ if the Bethe roots fulfill the conditions 
\begin{equation}\label{eq:constraint}
	P_+(\mu_j)P_0(\mu_j) = \prod_{l=2m+1}^M\frac{\sine{\mu_j+\lambda_l}{\eta}}{\sine{\mu_j-\lambda_l}{\eta}}\prod_{l=2m+1}^M\frac{\sine{\mu_j+\lambda_l}{0}}{\sine{\mu_j-\lambda_l}{0}} \neq 1
\end{equation}
for all paired Bethe roots $\mu_j = \lambda_{2j-1} = -\lambda_{2j}$, $j=1,\ldots,m$, and unpaired roots $\lambda_l$, $l=2m+1,\ldots,M$. 

If there are Bethe roots $\tilde{\mu}_j\in\{\mu_j\}_{j=1}^m$ that do not fulfill condition \eqref{eq:constraint}, we have $P_+(\tilde{\mu}_j)P_0(\tilde{\mu}_j)=1$. We define the index sets
\begin{subequations}
\begin{align}
	\tilde{J} &= \{k\in\{1,\ldots,2m\}|P_+(\lambda_k)P_0(\lambda_k)=1\quad\text{or}\quad P_-(\lambda_k)P_0(\lambda_k)=1\}\epc\\
	J &= \{k\in\{1,\ldots,2m\}|P_\pm(\lambda_k)P_0(\lambda_k)\neq 1\} \cup \{2m+1,\ldots,M\}\epc
\end{align}
\end{subequations}
and observe that $J\cap \tilde{J} = \emptyset$, $J\cup\tilde{J}=\{1,\ldots,M\}$, and if $2j-1 \in \tilde{J}$ then $2j\in\tilde{J}$ and vice versa. We choose the components of the vector $\vec{v}$ in the following way,
\begin{equation}
	v_j = \left\{ \begin{array}{lll}
	1 & \text{for} &  j \in J\epc \\
	0  &\text{for} &  j \in \tilde{J}
	 \epp
	\end{array} \right.
\end{equation}
The eigenvalue equations for the components $j\in \tilde{J}$ are as before. For $j\in J$ they read $(\tilde{B}\vec{v})_{2j-1} = \tilde{f}(\mu_j)$ and $(\tilde{B}\vec{v})_{2j} = \tilde{g}(\mu_j)$. The functions $\tilde{f}$ and $\tilde{g}$ are given by 
\begin{subequations}
\begin{align}
	\tilde{f}(\mu) &:= - \sum_{k=2m+1}^M\frac{\sine{0}{\eta}\sine{2\lambda_k}{0}}{\sine{\mu-\lambda_k}{\eta}\sine{\mu+\lambda_k}{0}}\prod_{\substack{l=2m+1\\l\neq k}}^M\frac{\sine{\lambda_k+\lambda_l}{0}}{\sine{\lambda_k-\lambda_l}{0}}\epc \\
	\tilde{g}(\mu) &:= - \sum_{k=2m+1}^M\frac{\sine{0}{\eta}\sine{2\lambda_k}{0}}{\sine{\mu+\lambda_k}{-\eta}\sine{\mu-\lambda_k}{0}}\prod_{\substack{l=2m+1\\l\neq k}}^M\frac{\sine{\lambda_k+\lambda_l}{0}}{\sine{\lambda_k-\lambda_l}{0}}\epp
\end{align}
\end{subequations}
 According to $f(\lambda)=g(\lambda)=0$ for all $\lambda\in\mathbb{C}$, we can write them as
 \begin{subequations}
\begin{align}
	\tilde{f}(\mu) &= \frac{\sine{0}{\eta}}{\sine{2\mu}{\eta}}\left(P_+(\mu)-P_0(\mu))^{-1}\right)\epc \\
	\tilde{g}(\mu) &= \frac{\sine{0}{\eta}}{\sine{2\mu}{-\eta}}\left(P_0(\mu)-P_-(\mu))^{-1}\right)\epp
\end{align}
\end{subequations}
 They have zeroes at $\mu=\tilde{\mu}_j$. 
 
In summary, we again found an eigenvector $\vec{v}\neq 0$ of the matrix $\tilde{B}$ with eigenvalue zero,
\begin{equation}
	\tilde{B}\vec{v} = 0 \qquad\text{where now}\quad v_j =\left\{ \begin{array}{ccl} 1 & \text{for} & j \in J\epc \\ 0 & \text{for} & j\in\tilde{J} \epp\end{array}\right.
\end{equation} 
The condition $\vec{v}\neq 0$ is equivalent to $\tilde{J}\neq\{1,\ldots,M\}$ which is always the case for $m<M/2$. In this case, we therefore have
 \begin{equation}
	\lim\nolimits_{\{\epsilon_j\to 0\}_{j=1}^m} \prod_{j=1}^m\frac{\sine{0}{\eta}}{\sine{2\epsilon_j}{0}}\det{}_{\!M}(B) = \det{}_{\!M}(\tilde{B})  = 0 
\end{equation}
if the set $\{\lambda_j\}_{j=1}^M$ satisfies Bethe equations. 

For $m=M/2$ we have $M/2$ pairs $\lambda_{2j-1}=-\lambda_{2j} \equiv \mu_j$ and $P_\pm(\mu_j) P_0(\mu_j)=1$ is always trivially fulfilled. Then $J=\emptyset$, $\tilde{J}=\{1,\ldots,M\}$, and $\vec{v}=0$. Thus, one cannot show that $\det_M(\tilde{B})=0$ this way. In fact, for the case of parity-invariant Bethe states $|\{\pm\mu_j\}_{j=1}^{M/2}\rangle$, it was previously shown \cite{XXZpaper} that 
\begin{equation}
	\lim\nolimits_{\{\epsilon_j\to 0\}_{j=1}^{M/2}} \prod_{j=1}^m\frac{\sine{0}{\eta}}{\sine{2\epsilon_j}{0}}\det{}_{\!M}(B) \neq 0\epp
\end{equation}
The explicit formulas are given in Eqs.~\eqref{eq:overlap_XXZ_offshell} and \eqref{eq:overlap_XXZ_offshell_odd} of Sec.~\ref{sec:pi_even}.

\subsubsection{States with one Bethe root at zero.}\label{sec:npi_odd}
Parity invariance constrains Bethe roots to occur in pairs or to be at zero and/or at $i\pi/2$. Let us consider the case when one Bethe root is at zero. If it is at $i\pi/2$ the logic of the proof is exactly the same and all calculations are similar and straightforward. We denote the set of off-shell parameters by $\{\lambda_j\}_{j=0}^{M-1}$ and set $\lambda_0=\epsilon_0$. We consider again $m$ pairs $\lambda_{2j-1}= \mu_j+\epsilon_j$, $\lambda_{2j} = -\mu_j+\epsilon_j$, $\mu_j\neq 0$, $j=1,\ldots, m$, with a complementary set of nonzero unpaired Bethe roots $\lambda_j \neq -\lambda_k$ for all $j,k = 2m+1, \ldots, M-1$ as in the previous section. The determinant then reads
\begin{align}
\det{}_{\!M}(\tilde{B}) &= \lim\nolimits_{\{\epsilon_j\to 0\}_{j=0}^m} \frac{\sine{0}{\eta/2}}{\sine{2\epsilon_0}{0}} \prod_{j=1}^m\frac{\sine{0}{\eta}}{\sine{2\epsilon_j}{0}}\det{}_{\!M}(B) \notag\\
	&= \det{}_{\!M}\begin{pmatrix} 
	\ast  & O_0 & \ldots &  \ldots & O_0 & B_{0,2m+1} & \ldots & B_{0, M-1} \\ 
\begin{array}{c}  \ast \\[-1.6ex] \ast \end{array} & \bar{B}_1  & O & \ldots & O & \begin{array}{l} \sss B_{1,2m+1} \\[-1.2ex] \sss B_{2,2m+1}\end{array} & \begin{array}{l} \sss\ldots \\[-1.2ex]\sss \ldots \end{array} & \begin{array}{l} \sss B_{1,M-1} \\[-1.2ex] \sss B_{2,M-1}\end{array} \\
\begin{array}{c}  \ast \\[-1.6ex] \ast \end{array}  & O & \bar{B}_2 & \ldots & O & \begin{array}{l} \sss B_{3,2m+1} \\[-1.2ex] \sss B_{4,2m+1}\end{array} & \begin{array}{l} \sss\ldots \\[-1.2ex]\sss \ldots \end{array} & \begin{array}{l} \sss B_{3,M-1} \\[-1.2ex] \sss B_{4,M-1}\end{array} \\
\vdots & \vdots & & \ddots & \vdots & \vdots & & \vdots \\
\begin{array}{c}  \ast \\[-1.6ex] \ast \end{array}  &O & O & \ldots & \bar{B}_m & \begin{array}{l} \sss B_{2m-1,2m+1} \\[-1.2ex] \sss B_{2m,2m+1}\end{array} & \begin{array}{l} \sss\ldots \\[-1.2ex]\sss \ldots \end{array} & \begin{array}{l} \sss B_{2m-1,M-1} \\[-1.2ex] \sss B_{2m,M-1}\end{array} \\
\begin{array}{c}  \ast \\[-1.6ex] \ast \end{array}  & O & O & \ldots & O & \begin{array}{l} \sss B_{2m+1,2m+1} \\[-1.2ex] \sss B_{2m+2,2m+1}\end{array} & \begin{array}{l} \sss\ldots \\[-1.2ex]\sss \ldots \end{array} & \begin{array}{l} \sss B_{2m+1,M-1} \\[-1.2ex] \sss B_{2m+2,M-1}\end{array} \\
\vdots &\vdots & & & \vdots & \vdots & \ddots & \vdots \\
\begin{array}{c}  \ast \\[-1.6ex] \ast \end{array}  &O & O & \ldots & O & \begin{array}{l} \sss B_{M-2,2m+1} \\[-1.2ex] \sss B_{M-1,2m+1}\end{array} & \begin{array}{l} \sss\ldots \\[-1.2ex]\sss \ldots \end{array} & \begin{array}{l} \sss B_{M-2,M-1} \\[-1.2ex] \sss B_{M-1,M-1}\end{array} \\
\end{pmatrix}\epc
\end{align}
where $O_0$ stands for $\begin{pmatrix} \ast & 0 \end{pmatrix}$, and all other blocks are the same as in Eq. \eqref{eq:block_matrix_even}.

Now we impose Bethe equations, {\it i.e.}~$b(\mu_j) = 1$ for $j=0,\ldots,m$ and $b_j = 1$ for $j=2m+1,\ldots,M-1$, where we set $\mu_0 = 0$ and identify index $M$ with index $0$. 
Then, considering the vector $\vec{v} = (0,1,1\ldots,1)^t$, we obtain 
\begin{equation}
(\tilde{B}\vec{v})_{0}  =  
	 - \sum_{k=2m+1}^{M-1}\frac{\sine{0}{\eta}\sine{2\lambda_k}{0}}{\sine{\lambda_k}{-\eta}\sine{-\lambda_k}{0}}\prod_{\substack{l=2m+1\\l\neq k}}^{M-1}\frac{\sine{\lambda_k+\lambda_l}{0}}{\sine{\lambda_k-\lambda_l}{0}} 
	=\left(1 -  \prod_{l=2m+1}^{M-1}\frac{\sine{\lambda_l}{\eta}}{\sine{\lambda_l}{-\eta}} \right) = 0 \epp
\end{equation}
In the last step we used $g(\mu= 0) = 0$ from Eq.~\eqref{eq:g_function} with $M$ replaced by $M-1$. Under the condition that $2m \neq M-1$ we can further show that $(\tilde{B}\vec{v})_{j} = 0$ for all other components, analogously to the case of even number of particles in subsection \ref{sec:npi_even}.

Hence, it is shown that both $|N\rangle$ and $|AN\rangle$ have zero overlap with all non-parity-invariant on-shell Bethe states. Therefore, the overlap with the zero-momentum N\'eel state $|\Psi_0\rangle = (|N\rangle+|AN\rangle)/\sqrt{2}$ is zero, too.

\section{Scaling to the Lieb-Liniger Bose gas for an odd number of bosons}\label{sec:scaling_LL}
Before we consider the scaling limit of the overlap $\langle\Psi|\{\lambda_j\}_{j=1}^M\rangle$ to Lieb-Liniger \cite{GaudinBOOK, 1987_Golzer, 2007_Seel, Pozsgay_JStatMech_P11017} ($N\epsilon^2 = cL$, $\eta = i\pi - i\epsilon$, $\lambda_j \to \epsilon\lambda_j/c$ for all finite Bethe roots, and eventually $\epsilon\to 0$), we have to ensure that there are only finitely many downturned spins in the initial state $|\Psi\rangle$ because then the overlap is nonzero only for Bethe states with finite $M$ and the scaling limit to LL is applicable. The norm of a parity-invariant on-shell Bethe state with an odd number $M=2m+1$ of downturned spins is given by [see Eq.~\eqref{eq:norm_Bethe_state}]
\begin{multline}
\| \{ \pm\lambda_{j} \}_{j=1}^{m} \cup \{\lambda_0=0\}\| = \sine{0}{\eta}^{m+\frac{1}{2}} \left[\prod_{j=1}^m \frac{\sqrt{ \sine{2\lambda_j}{\eta}  \sine{2\lambda_j}{-\eta} }}{\sine{2\lambda_j}{0}}\right] \\ 
 \times\left[\prod_{j=1}^m \frac{\sine{\lambda_j}{\eta}  \sine{\lambda_j}{-\eta} }{\sine{\lambda_j}{0}^2}\right]  \prod_{\substack{j>k=1\\ \sigma = \pm}}^m\frac{\sine{\lambda_j - \sigma \lambda_k}{\eta}  \sine{\lambda_j + \sigma \lambda_k}{-\eta} }{\sine{\lambda_j + \sigma \lambda_k}{0}^2} \sqrt{\det{}_{\!2m+1} (G_{jk}) }\epp
\end{multline}
To derive a determinant expression for $\langle\Psi | \{\lambda_j\}_{j=1}^M\rangle$ we follow exactly the logic of Ref.~\cite{XXZ_LL_proof}. As initial states we consider $2n$-fold $q$-raised N\'eel states 
\begin{align}
|N^{(2n)}\rangle &= (S_q^+\tilde{S}_q^+)^{n}|N\rangle\epc\quad\text{where}\\
	S_q^+ &= \sum_{n=1}^N \left[\prod_{j=1}^{n-1} q^{\sigma_j^z/2}\right] \sigma_n^+ \left[\prod_{j=n+1}^N q^{-\sigma_j^z/2}\right]\epc\quad
	\tilde{S}_q^+ = \sum_{n=1}^N \left[\prod_{j=1}^{n-1} q^{-\sigma_j^z/2}\right] \sigma_n^+ \left[\prod_{j=n+1}^N q^{\sigma_j^z/2}\right]\notag
\end{align}
and $q$ is related to the anisotropy via $\Delta = (q+q^{-1})/2$. In the LL scaling limit, these $q$-deformed operators act on the N\'eel states as global $SU(2)$ operators and the state scales to the (unnormalized) BEC-like state \cite{LLpaper} of the Lieb-Liniger Bose gas. If we set $N/2=2n+2m+1$ this leads to an odd number of downturned spins in the resulting initial $|N^{(2n)}\rangle$ state, which directly corresponds to an odd number of bosons in the BEC state. 

\subsection{Generalization of the overlap formula to q-raised N\'eel states}\label{sec:q-raised}
We make use of overlap formula \eqref{eq:overlap_XXZ_offshell_odd} for an odd number of downturned spins with $\lambda_0=0$. We consider the overlap of the $2n$-fold $q$-raised N\'eel state with a parity-invariant (off-shell) Bethe state in the magnetization sector $2n = N/2-(2m+1)$. The global symmetry operators that are needed to construct $q$-raised N\'eel states are obtained by sending spectral parameters to infinity and taking the proper normalization into account \cite{XXZ_LL_proof}. 

As in the even particle case \cite{XXZ_LL_proof} we split the prefactor $\gamma$ of the determinant formula into two parts,
\begin{equation}
	\gamma = \gamma_\infty\hat{\gamma}\epp
\end{equation}
We again have $\gamma_\infty = \left(\frac{-1}{4\sine{0}{\eta}^2}\right)^{n}$. The other part of $\gamma$ is determined by
\begin{multline}
\frac{\hat\gamma}{\| \{ \pm\lambda_{j} \}_{j=1}^{m} \cup \{\lambda_0=0\} \|} = \frac{\sinh(\frac{\eta}{2})}{\sinh(\eta)}
\left[\prod_{j=1}^{m}\frac{\sqrt{\tanh(\lambda_j+\eta/2)\tanh(\lambda_j-\eta/2)}}{2\sinh(2\lambda_j)}\right]\\ \times\frac{1}{\sqrt{\sinh^{2m+1}(\eta)\det{}_{\!2m+1}(G_{jk}) } }\epp
\end{multline}

Within the limit of some spectral parameters to infinity the determinant can be simplified. Following the reasoning of Ref.~\cite{XXZ_LL_proof} the entire upper right $(m+1)\times n$ block is zero. Furthermore, the lower right $n\times n$ block becomes a triangular matrix and together with the factor $\gamma_\infty$ the determinant of this block becomes again $[2n]_q!$ (as in the even particle case). Then we plug in the on-shell condition and all $\mathfrak{A}$-terms in the upper left $(m+1)\times(m+1)$ block vanish. In summary, 
\begin{subequations}\label{eq:new_overlap_odd}
\begin{align}\label{eq:new_overlap_a}
	 \frac{\langle N^{(2n)} |\{\pm\lambda_j\}_{j=1}^{m}\cup \{\lambda_0=0\} \rangle}{\| |N^{(2n)}\rangle \| \|\{\pm\lambda_j\}_{j=1}^m \cup \{0\} \|} &=  \frac{[2n]_q!}{\| |N^{(2n)}\rangle \|} \frac{\hat{\gamma}\det{}_{\!m+1}(\hat{G}_{jk}^{+})}{\|\{\pm\lambda_j\}_{j=1}^m \cup \{0\} \|} \notag \\
&= \frac{[2n]_q!}{\| |N^{(2n)}\rangle\|} \frac{\sinh(\frac{\eta}{2})}{\sinh(\eta)}\left[\prod_{j=1}^{m}\frac{\sqrt{\tanh(\lambda_j+\frac{\eta}{2}) \tanh(\lambda_j-\frac{\eta}{2})}}{2\sinh(2\lambda_j)}\right]
\notag\\ 
&\quad\times\frac{\det{}_{\!m+1}(\hat{G}_{jk}^{+})}{\sqrt{\det{}_{\!2m+1}(\hat{G}_{jk})}}\epc
\end{align}
where
\begin{align}
\hat{G}_{jk}^+ &= \delta_{jk}\left(N\sine{0}{\eta}K(j)-\sum_{l=0}^{m}\sine{0}{\eta}K^+(j,l)\right) + \sine{0}{\eta}K^+(j,k)\epc \quad 0\leq j,k\leq m\epc \\ 
\hat{G}_{jk} &= \delta_{jk}\left(N\sine{0}{\eta}K_{\eta/2}(\lambda_j)-\sum_{l=0}^{2m}\sine{0}{\eta}K_\eta(\lambda_j-\lambda_l)\right) + \sine{0}{\eta}K_\eta(\lambda_j-\lambda_k)\epc \quad 0\leq j,k\leq 2m\epp
\end{align}
\end{subequations}
Note again that $K^+(j,k)$ is defined differently for $j,k = 0$ and for $1\leq j,k \leq m$ (by a factor $1/2$). The function $K_\eta$ is defined as usual, $K_{\eta}(\lambda) = \frac{\sinh(2\eta)}{\sinh(\lambda+\eta)\sinh(\lambda-\eta)}$. $\| |N^{(2n)}\rangle\|$ is the norm of the $2n$-fold $q$-raised N\'eel state which, in the limit $q\to -1$, can be calculated explicitly \cite{XXZ_LL_proof},
\begin{equation}
\left\| |N^{(2n)}\rangle \right\| =  (2n)! \sqrt{\begin{pmatrix}
N/2 \\
2m + 1
\end{pmatrix}} = \frac{(2n)!}{\sqrt{(2m +1)!}} \left(\frac{N}{2}\right)^{m+1/2} \left(1+ \mathcal{O}(1/N) \right)\epp
\end{equation}
We shall need this result to evaluate the prefactor of the determinant formula in the scaling limit to LL exactly. 

\subsection{Overlap of a LL Bethe state with the BEC state with an odd number of bosons}\label{sec:LL_odd}
We repeat the steps of Ref.~\cite{XXZ_LL_proof} for the scaling limit to Lieb-Liniger. There is a one-to-one correspondence \cite{1987_Golzer} between XXZ and Lieb-Liniger Bethe states as well as between their norm formulas. The matrices $G_{jk}^{+,odd}$ in Eq.~\eqref{eq:overlap_XXZ_offshell_odd} for finite $M'$ turn into the corresponding Lieb-Liniger matrices \cite{LLpaper}.
 
Since $\sinh(\frac{\eta}{2})/\sinh(\eta) \sim 1/\epsilon$ and due to the corrective factor $2^{m+1/2}$ (for an explanation of this factor see Ref.~\cite{XXZ_LL_proof}), we obtain for the overlaps with the BEC state, $\langle x | BEC\rangle = L^{-N_{LL}/2}$ (with $N_{LL} = 2m+1$, $m$ being the number of LL rapidities with positive real part)
\begin{subequations}
\begin{equation}
	\frac{\langle BEC | \{\pm\lambda_j\}_{j=1}^{m}\cup\{0\}\rangle}{\| \{\pm\lambda_j\}_{j=1}^{m}\cup\{\lambda_0=0\}\|} = \frac{\ (cL)^{-N_{LL}/2}\sqrt{N_{LL}!}}{\displaystyle \prod_{j=1}^{m} \frac{\lambda_j}{c}\sqrt{\frac{\lambda_j^2}{c^2}+\frac{1}{4}} }\  \frac{2\det{}_{\!m+1}\left(G_{jk}^{(+,LL,odd)}\right)}{\sqrt{\det{}_{\!2m+1}G_{jk}^{LL}}}
\end{equation}
with
\begin{align}
G_{jk}^{(+,LL,odd)} &= \delta_{jk}\left(\frac{cL}{1+\delta_{j0}} + \sum_{l=0}^{m} K_{LL}(j,l)\right) - K_{LL}(j,k)\epc \quad j,k=0,\ldots,m\epc \\ 
G_{jk}^{LL} &= \delta_{jk}\left(cL+\sum_{l=0}^{2m}K_{LL}(\lambda_j-\lambda_l)\right) - K_{LL}(\lambda_j-\lambda_k)\epc \quad j,k=0,\ldots,2m\epc\\[2ex]
K_{LL}(j,k) &= \left\{ \begin{array}{r@{\hspace{4ex}}l} \left(K_{LL}(\lambda_j-\lambda_k) + K_{LL}(\lambda_j+\lambda_k)\right)\epc & 1\leq j,k \leq m \epc \\  \frac{1}{2} \left(K_{LL}(\lambda_j-\lambda_k) + K_{LL}(\lambda_j+\lambda_k)\right) \epc & \text{otherwise}\epc \end{array} \right.\\
K_{LL}(\lambda) &= \frac{2c^2}{\lambda^2+c^2}\epp
\end{align}
\end{subequations}
We cross-checked this result analytically for $N_{LL}=3$ and numerically up to $N_{LL}=9$. This completes the result of Ref.~\cite{XXZ_LL_proof} where the formula for overlaps of parity-invariant LL Bethe states with  BEC states with even numbers of bosons \cite{LLpaper} was proven. The results of Sec.~\ref{sec:npi} for non-parity-invariant XXZ Bethe states are valid in the scaling limit to LL which proves that overlaps of the BEC state with non-parity-invariant LL Bethe states are zero.

\section{Conclusion}
In this paper we presented a proof that the overlap of the N\'eel state with a non-parity-invariant on-shell Bethe state vanishes. We further gave a formula for the overlap with parity-invariant Bethe states with an odd number of downturned spins. These results of subsections \ref{sec:pi_odd}, \ref{sec:npi_even}, and \ref{sec:npi_odd} complete the answer to the question: what is the overlap of the N\'eel state with XXZ Bethe states? It is now proven that it is zero for any non-parity-invariant on-shell state, and it is nonzero for parity-invariant (off-shell) states. The latter can be expressed by ``Gaudin-like'' determinants for both cases, even and odd number of down spins. This therefore opens the door to a quench action analysis of the quench starting from the N\'eel state. Our results thus pave the way towards a comparison with GGE predictions of this quench, as obtained in \cite{2013_Pozsgay_JSTAT, Fagotti_1311.5216, 2013_Fagotti}, which would be particularly interesting.

The scaling to the Lieb-Linger model and the derivation of the corresponding determinant expressions for even and odd numbers of bosons is straightforward (for the former see Ref.~\cite{XXZ_LL_proof}, for the latter see Sec.~\ref{sec:LL_odd}). To summarize, the overlaps of the BEC state with Bethe states of the Lieb-Liniger Bose gas (repulsive and attractive) are now known for all possible cases. They vanish for non-parity-invariant states and are, in general, nonzero for parity-invariant ones. The latter are given by determinants of Gaudin type, which have slightly different structures for even and an odd numbers of bosons. These results can be used e.g.~in the context of 1D Kardar-Parisi-Zhang equation as in \cite{Calabrese_1402.1278}, which can be mapped to the attractive Lieb-Liniger Bose gas.

\section*{Acknowledgements}
We would like to thank Frank G{\"o}hmann and Pasquale Calabrese for useful discussions. We acknowledge financial support from the Foundation for Fundamental Research on Matter (FOM) and the Netherlands Organisation for Scientific Research (NWO).

\appendix

\section{Derivation of the overlap formula for an odd number of down spins}\label{sec:odd_proof}
We start from equation \eqref{eq:overlap_S3} for the overlap of an off-shell Bethe state $|\{\tilde{\lambda}_j\}_{j=1}^M\rangle$ again with $N=2M$, but now $M$ odd.  
We set $M'=(M-1)/2$ and further $\tilde\lambda_j = \lambda_j + \epsilon_j$ for $j=1,\ldots,M'$, $\tilde\lambda_j = -\lambda_{j-M'} + \epsilon_{j-M'}$ for $j=M'+1,\ldots,M-1$, and finally $\tilde{\lambda}_M \equiv \lambda_0 + \epsilon_0$. We first consider $\lambda_0=0$. The case $\lambda_0 = i\pi/2$ is discussed at the end of this section.
By multiplying the prefactor and the inverse of the determinant 
with $\prod_{j=0}^{M'} \delta_j $, where $\delta_j = \frac{\sinh (2 \epsilon_j)}{\sinh(\eta)}$, we get regular expressions in the limits $\epsilon_j\to 0$,  
\begin{align}\label{eq:gamma_App}
	&\gamma = \frac{\sine{0}{\eta/2} }{\sine{0}{\eta}} \left[\prod_{j=1}^{M'}\frac{\sine{\lambda_j}{\eta/2}\sine{\lambda_j}{-\eta/2}}{\sine{2\lambda_j}{0}^2}\right] \left[\prod_{j=1}^{M'}  \frac{\sine{\lambda_j}{\eta}\sine{\lambda_j}{-\eta}}{\sine{\lambda_j}{0}^2}\right] 
	\left[\prod_{\substack{j>k=1\\ \ \sigma=\pm}}^{M'}\frac{\sine{\lambda_j+\sigma\lambda_k}{\eta}\sine{\lambda_j+\sigma\lambda_k}{-\eta}}{\sine{\lambda_j+\sigma\lambda_k}{0}^2}\right]\epc \\
&\det{}_\text{reg} = \lim_{\delta_0 \to 0}\lim\nolimits_{\{\delta_j \to 0\}_{j=1}^{M'}} \left\{\left[\prod_{j=0}^{M'} \delta_j \right]^{-1} \det{}_{\!M}(\delta_{jk}+U_{jk})\right\}\epp \label{eq:Dreg}
\end{align}
We order the rows and columns of the matrix in such a way that the first one corresponds to $\lambda_0$, and that pairs of rapidities $\lambda_j=-\lambda_k$ always correspond to neighboured rows and columns. We label them by indices $0,1,\ldots,2M'$. Column zero of the matrix is up to first order in $\delta_0$ given by
\begin{subequations}
\begin{align}
1 + U_{00} &=  \frac{\delta_0}{2} \left(2M\sine{0}{\eta} K_{\frac{\eta}{2}} (\lambda_0)  - \sum_{l=1}^{M'} \sine{0}{\eta} K^+(\lambda_0, \lambda_l) \right)\\ \label{eq:U2jm1}
	U_{2j-1,0} &= \delta_{0}  \frac{\sine{0}{\eta}^2}{\sine{\lambda_j+\lambda_0}{0} \sine{\lambda_j-\lambda_0}{\eta} }\mathfrak{a}_0 = \delta_{0}  \frac{\sine{2\lambda_0}{\eta}\sine{0}{\eta}}{\sine{\lambda_j+\lambda_0}{0} \sine{\lambda_j-\lambda_0}{\eta} }\mathfrak{a}_0 = \delta_{0} \tilde{U}_{2j-1,0}\epc\\
	U_{2j,0} &= \delta_{0}  \frac{\sine{0}{\eta}^2}{\sine{\lambda_j-\lambda_0}{0} \sine{\lambda_j+\lambda_0}{-\eta} }\mathfrak{a}_0 = \delta_{0}  \frac{\sine{2\lambda_0}{\eta}\sine{0}{\eta}}{\sine{\lambda_j-\lambda_0}{0} \sine{\lambda_j+\lambda_0}{-\eta} }\mathfrak{a}_0 = \delta_{0} \tilde{U}_{j0}\epp
\end{align}
\end{subequations}
The second step is only allowed when $\lambda_0 = 0$. For $\lambda_0=i\pi/2$ it means that there would be an additional minus sign. 

Now we insert the parity invariance condition for all other $2M'=M-1$ rapidities. We can do exactly the same procedure as for the even case \cite{XXZpaper}. We redefine the factors $\mathfrak{a}_k$ as
\begin{subequations}
\begin{align}\label{eq:func_a_tilde_App}
  \mathfrak{a}_k &= \frac{\sinh(\lambda_k-\lambda_0-\eta)}{\sinh(\lambda_k-\lambda_0+\eta)} \left[\prod_{\substack{l=1\\ \ \sigma=\pm}}^{M'}\frac{\sine{\lambda_k-\sigma\lambda_l}{-\eta}}{\sine{\lambda_k-\sigma\lambda_l}{\eta}}\right]\left(\frac{\sine{\lambda_k}{\eta/2}}{\sine{\lambda_k}{-\eta/2}}\right)^{2M}\epc \quad k =1,\ldots,M'\epc \\
 \mathfrak{a}_0 &= -1\epp
\end{align}
\end{subequations}
The first row of the matrix, row zero, is then given by
\begin{subequations}
\begin{align}
U_{0,2k-1} &=  \delta_k\frac{\sine{2\lambda_k}{\eta}\sine{0}{\eta}}{\sine{\lambda_k-\lambda_0}{0}\sine{\lambda_0-\lambda_k}{\eta}}\frac{\sine{\lambda_k-\lambda_0}{\eta}}{\sine{\lambda_k+\lambda_0}{\eta}}\mathfrak{a}_k = \delta_k\frac{\sine{2\lambda_k}{\eta}\sine{0}{\eta}}{\sine{\lambda_0-\lambda_k}{0}\sine{\lambda_0+\lambda_k}{-\eta}}\mathfrak{a}_k = \delta_k  \tilde{U}_{0,2k-1}  \epc \\
U_{0,2k} &= \delta_k\frac{\sine{2\lambda_k}{-\eta}\sine{0}{\eta}}{\sine{\lambda_k+\lambda_0}{0}\sine{\lambda_k+\lambda_0}{\eta}}\frac{\sine{\lambda_k+\lambda_0}{-\eta}}{\sine{\lambda_k-\lambda_0}{-\eta}}\mathfrak{a}_k^{-1} = \delta_k\frac{\sine{2\lambda_k}{-\eta}\sine{0}{\eta}}{\sine{\lambda_0+\lambda_k}{0}\sine{\lambda_k-\lambda_0}{\eta}}\mathfrak{a}_k^{-1} = \delta_k  \tilde{U}_{0,2k}\epp
\end{align}
\end{subequations}
These steps are allowed for both $\lambda_0=0$ and $\lambda_0=i\pi/2$. We further define $\alpha_k = \sqrt{-\frac{\sine{2\lambda_k}{\eta}}{\sine{2\lambda_k}{-\eta}}\mathfrak{a}_k}$ as in Ref.~\cite{XXZpaper}, $\alpha_0=i$ and we multiply the $M\times M$ matrix $\mathds{1}+U$ from the left and from the right respectively with the diagonal matrices
\begin{equation}
	\text{diag}_{M}\left(\alpha_0^{-1}, \alpha_1,\alpha_1^{-1},\ldots,\alpha_{M'},\alpha_{M'}^{-1}\right)\epc \quad \text{diag}_{M}\left(\alpha_0,\alpha_1^{-1},\alpha_1,\ldots,\alpha_{M'}^{-1},\alpha_{M'}\right)\epp 
\end{equation}
The structure of the matrix becomes {\scriptsize \vspace{2ex}
\begin{equation*}\label{eq:matrix_structure_odd}	
	\left(\begin{array}{cccc} 	
	\delta_0 \tilde{U}_{00} & \delta_1 \alpha_0^{-1} \left[\!\begin{array}{cc} \tilde{U}_{0,1} \alpha_1^{-1} &  \tilde{U}_{0,2} \alpha_1 \end{array}\!\right]  & \delta_2 \alpha_0^{-1}\left[\!\begin{array}{cc} \tilde{U}_{0,3} \alpha_2^{-1} &  \tilde{U}_{M,4} \alpha_2 \end{array}\!\right]  & \dots  \\[4ex]
	 \delta_0 \alpha_0\left[\!\!\begin{array}{c} \alpha_1 \tilde{U}_{1,0}  \\ \alpha_1^{-1} \tilde{U}_{2,0} \end{array}\!\!\right] &
	\left[\!\!\!\begin{array}{ll} 1-\delta_1\frac{\sine{2\lambda_1}{-\eta}}{\sine{2\lambda_1}{0}}\alpha_1^2 & 1+\delta_1\mathfrak{b}_1^{-}\alpha_1^2\\ 1+\delta_1\mathfrak{b}_1^{+}\alpha_1^{-2} & 1-\delta_1\frac{\sine{2\lambda_1}{\eta}}{\sine{2\lambda_1}{0}}\alpha_1^{-2} \end{array}\!\!\right] & \delta_2\left[\!\!\begin{array}{cc}a_{12}&b_{12}\\ c_{12}&d_{12}\end{array}\!\!\right] & \dots	\\[6ex]
	\delta_0 \alpha_0 \left[\!\!\begin{array}{c}  \alpha_2 \tilde{U}_{3,0}  \\ \alpha_2^{-1} \tilde{U}_{4,0} \end{array}\!\!\right] &
	\delta_1\left[\!\!\begin{array}{cc}a_{21} & b_{21}\\ c_{21} & d_{21}\end{array}\!\!\right] & \left[\!\!\!\begin{array}{ll} 1-\delta_2\frac{\sine{2\lambda_2}{-\eta}}{\sine{2\lambda_2}{0}}\alpha_2^2 & 1+\delta_2\mathfrak{b}_2^{-}\alpha_2^2\\ 1+\delta_2\mathfrak{b}_2^{+}\alpha_2^{-2} & 1-\delta_2\frac{\sine{2\lambda_2}{\eta}}{\sine{2\lambda_2}{0}}\alpha_2^{-2} \end{array}\!\!\right] & \dots \\
	\vdots & \vdots & \vdots & \ddots 
\end{array}\right)\epc 
\end{equation*}}
where the elements $a_{jk}$, $b_{jk}$, $c_{jk}$, and $d_{jk}$, $j,k=1,\ldots M'$ are the same as in the even case. Analogously, the determinant can be simplified by replacing column $2k-1$ by the difference of columns $2k-1$ and $2k$ for all $k=1,\ldots, M'$ and afterwards by replacing row $2j-1$ by the difference of rows $2j-1$ and $2j$ for all $j=1,\ldots, M'$ as in Ref.~\cite{XXZpaper} which leads up to first order to
\begin{multline}\label{eq:matrix_structure2}
	\det{}_{\!M}  \left(\begin{array}{cccc} 
	 \delta_0 \tilde{U}_{00} & \left[\!\!\begin{array}{cc} \delta_1 e_{01} & 0 \end{array}\!\!\right] & \left[\!\!\begin{array}{cc} \delta_2 e_{02} & 0 \end{array}\!\!\right] & \dots \\[1ex]
	  \left[\!\!\begin{array}{c} \delta_0 e_{10} \\ 0 \end{array}\!\!\right] &
	\left[\!\!\begin{array}{c@{\hspace{0.9ex}}c} \delta_1 D_{1}  & 0 \\ 0 & 1 \end{array}\!\!\right] &
	\left[\!\!\begin{array}{c@{\hspace{0.9ex}}c} \delta_2 e_{12} & 0 \\ 0 & 0 \end{array}\!\!\right] & \dots\\[3ex]
	\left[\!\!\begin{array}{c} \delta_0 e_{20} \\ 0 \end{array}\!\!\right] &
	\left[\!\!\begin{array}{c@{\hspace{0.9ex}}c} \delta_1 e_{21} & 0 \\ 0 & 0 \end{array}\!\!\right] &
	\left[\!\!\begin{array}{c@{\hspace{0.9ex}}c} \delta_2 D_{2}  & 0 \\ 0 & 1 \end{array}\!\!\right] & \\
	\vdots & \vdots & & \ddots 
	\end{array}\right) \\[2ex]
	=  \left[\prod_{k=0}^{M'}\delta_k\right]\det{}_{\!M'+1}
	\hspace{-0.5ex}\left[\begin{array}{cccc} 
	\tilde{U}_{00} & e_{01} & e_{02} & \dots\\
	e_{10}  & D_{1}  & e_{12} & \dots \\
	 e_{2M} & e_{21} & D_{2}  & \\[-0.7ex]
	\vdots & \vdots & & \ddots 
	\end{array}\right]\epp
\end{multline}
The new matrix elements are ($\lambda_0 = 0$ and $\lambda_0 = i\pi/2$)
\begin{subequations}
\begin{align}
e_{jk} = & \sine{0}{\eta}  K^+(\lambda_j - \lambda_k) + f_{jk} \qquad \text{for}\quad  j,k =1,\ldots, M'\epc\\
e_{j0} =& \frac{1}{2}\sine{0}{\eta} K^+(\lambda_j, \lambda_0) + f_{j0}\epc \\
e_{0k} =& \frac{1}{2}\sine{0}{\eta} K^+(\lambda_0, \lambda_k) + f_{0k} \epc\\
\tilde{U}_{00}   = &   \frac{1}{2} \left( 2M\sine{0}{\eta}  K_{\frac{\eta}{2}} (\lambda_0)  - \sum_{l=1}^{M'} \sine{0}{\eta} K^+(\lambda_0, \lambda_l) \right) \epc
\end{align}
\end{subequations}
where we used $K_{\eta}(\lambda_j ) = \frac{1}{2} K^+(\lambda_j ,\lambda_0)$. If $\lambda_0 = i\pi/2$ the formula for the overlap is the same and its derivation straightforward. The only difference is a minus sign in Eq.~\eqref{eq:U2jm1}, which cancels at the end when we express the corresponding matrix elements by the function $K^+(\lambda_j,\lambda_0)$. This eventually leads to Eqs.~\eqref{eq:overlap_XXZ_offshell_odd}.

\end{document}